\documentclass[preprint,superscriptaddress,longbibliography]{revtex4-1}

\usepackage[utf8]{inputenc}
\usepackage[pdftex]{graphicx}
\usepackage{amsmath}
\usepackage[amssymb,Gray]{SIunits}
\usepackage{bm}
\usepackage[all]{xy}

\newcommand{\abs}[1]{\ensuremath{\left\vert #1 \right\vert}}
\newcommand{\D}{\mathrm{d}}

\renewcommand{\vec}[1]{\mathbf{#1}}
\newcommand{\bleq}{\ensuremath{\mathrel{\phantom{=}}}}
\newcommand{\nnl}{\nonumber\\}

\newcommand{\com}{c.m.}

\newcommand{\pp}{\mathcal{P}}
\newcommand{\ppv}{\bm{\mathcal{P}}}
\newcommand{\ph}{\mathcal{H}}
\newcommand{\pjv}{\bm{\mathcal{J}}}
\newcommand{\pkv}{\bm{\mathcal{K}}}
\newcommand{\pj}{\mathcal{J}}
\newcommand{\pk}{\mathcal{K}}

\begin{document}


\title{Quantum mechanics for non-inertial observers}


%
\author{Marko Toro\v{s}}
\email[]{marko.toros@ts.infn.it}
\author{Andr\'e Gro{\ss}ardt}
\email[]{andre.grossardt@ts.infn.it}
\author{Angelo Bassi}
\email[]{bassi@ts.infn.it}
\affiliation{Department of Physics, University of Trieste, Strada Costiera 11, 34151 Miramare-Trieste, Italy}
\affiliation{Istituto Nazionale di Fisica Nucleare, Sezione di Trieste, Via Valerio 2, 34127 Trieste, Italy}


\date{\today}

\begin{abstract}
A recent analysis by Pikovski et al.~[Nat. Phys. 11, 668 (2015)] has triggered interest in
the question of how to include relativistic corrections in the quantum dynamics governing
many-particle systems in a gravitational field. Here we show how the
center-of-mass motion of a quantum system subject to gravity can be derived
more rigorously, addressing the ambiguous definition of relativistic center-of-mass coordinates.
We further demonstrate that, contrary to the prediction by Pikovski et al., external forces
play a crucial role in the relativistic coupling of internal and external degrees of freedom,
resulting in a complete cancellation of the alleged coupling in Earth-bound laboratories for systems supported against gravity
by an external force. We conclude that the proposed decoherence effect is an effect of
relative acceleration between quantum system and measurement device, rather than a universal
effect in gravitational fields.
\end{abstract}


\maketitle



How does a quantum particle move in a gravitational field?
Thus far, only the nonrelativistic case of a particle in a Newtonian, homogeneous gravitational
potential has been experimentally tested~\cite{Colella:1975}. 
The relativistic description of the dynamics of interacting quantum particles, on the other hand,
is believed to require quantum field theory~\cite{Currie:1963}, since relativistic effects
must include particle creation and annihilation.
Strictly speaking, general relativistic corrections to the evolution
of quantum matter should, therefore, be described within the framework of quantum field theory in
curved spacetime~\cite{Wald:1994,Brunetti:2015}.

Nonetheless, it is instructive and of relevance for experiments~\cite{kovachy2015quantum,Arndt:2014,Eibenberger:2013,Zoest:2010,Altschul:2015,amelino2013quantum,Belenchia:2016}
to discuss relativistic effects in terms of higher order
corrections to the nonrelativistic Schrödinger equation: a model applicable to quantum
systems for which the number of particles is conserved, to good approximation.
These corrections can, for instance, be obtained as higher orders in $1/c^2$ of the
$c \to \infty$ expansion for a classical Klein-Gordon field in curved
spacetime~\cite{Kiefer:1991,Laemmerzahl:1995}.
This result, however, must be considered as a \emph{single particle} evolution equation.
A more difficult problem, within this scheme, is how to analyze a system of many interacting
particles. This is a necessary step towards correctly understanding how relativistic corrections
affect the motion of atoms or molecules. The first problem one faces, is how to define a
center-of-mass wave-function and how to derive its dynamics.

It is tempting to extend the one-particle dynamics to the center of mass (\com) of a composite system, simply
replacing the one-particle rest mass $m$ by $M + H_\text{rel}^{(0)}/c^2$, $M$ being the sum of all
rest masses and $H_\text{rel}^{(0)}$ the internal Hamiltonian to nonrelativistic order. One then
obtains, up to order $1/c^2$, a coupling of internal and \com\ degrees of freedom proportional to
$(M c^2 - T^{(0)} + U^{(0)}) H_\text{rel}^{(0)}/c^2$, where $T^{(0)}$ and $U^{(0)}$ are the nonrelativistic
\com\ kinetic and gravitational potential energy, respectively. Based on this coupling,
decoherence of the \com\ wave-function due to an external, homogeneous
gravitational field was predicted by Pikovski et~al.~\cite{Pikovski:2015},
followed by a vivid debate~\cite{Adler:2016,Bonder:2016,Bonder:2015a,Diosi:2015,Pang:2016}.
One immediate question which, to date, has not been raised, is the compatibility of
this scheme with the well-known problems in defining relativistic \com\
variables~\footnote{We refer to the position and momentum of the mass center as \com\
\emph{coordinates}, and to the collective set of \com\ and relative coordinates as
\com\ \emph{variables}.}:
no definition exists in which the \com\ is (i) uniformly moving in absence of external forces,
(ii) positions and momenta obey the canonical commutation relations, and (iii) the \com\ is
frame independent~\cite{Pryce:1948}.
A perturbative procedure to construct canonical \com\ variables, in which the dynamics in
flat spacetime take the single particle form, was presented by Krajcik and Foldy~\cite{Krajcik:1974}.

In this letter, we show how to generalize this prescription to accelerated observers, or
to a homogeneous gravitational field by virtue of the equivalence principle.
Thereby, a precise definition of the coordinates in which the alleged coupling of \com\ and internal degrees of freedom holds can be given.

\begin{figure}
\includegraphics[scale=.27]{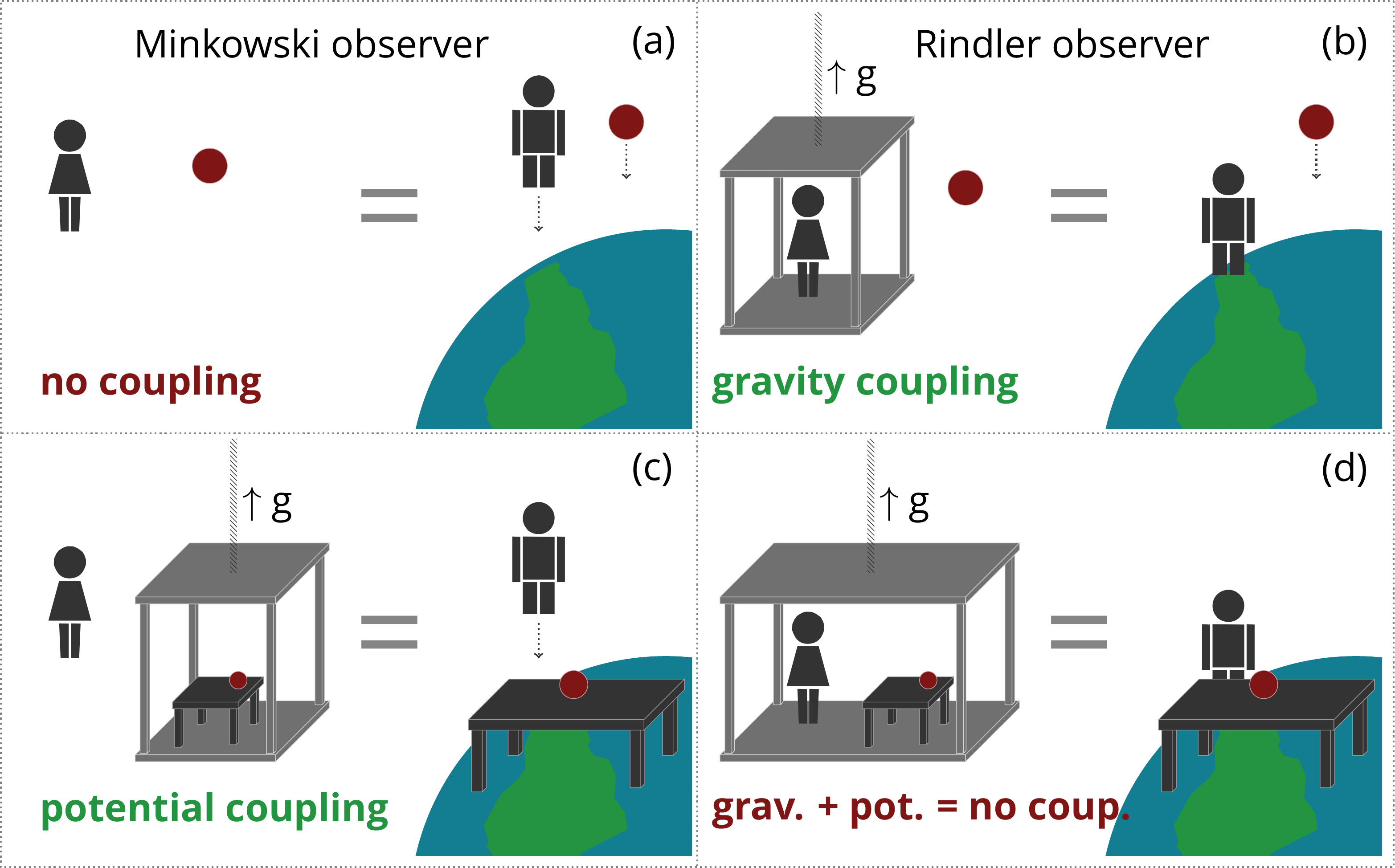} 
\caption{Overview of the different observer/particle constellations:
(a) inertial (Minkowski) observer with free particle;
(b) accelerated (Rindler) observer with free particle;
(c) inertial observer with accelerated particle;
(d) accelerated observer with accelerated particle.
The equality signs denote that, according to the equivalence principle,
inertial and accelerated motion correspond to being in free fall and at rest with
respect to the Earth, respectively.
In situation (b) a coupling of \com\ and internal degrees of freedom occurs due to
acceleration. In (c) the opposite coupling occurs due to the accelerating external
potential. In (d) both effects cancel each other.
\label{fig:summary}}
\end{figure}

Our analysis shows that, in this precise context, the \com\ motion couples to the
internal dynamics, both for a particle held at constant height
in the gravitational potential as seen by a free falling observer and for a free falling particle
as seen by an accelerated (e.\,g. Earth bound) observer (see Fig.~\ref{fig:summary}\,b,\,c).
Yet, no coupling of the \com\ to the internal Hamiltonian occurs in situations where both
the particle and the observer are accelerated (see Fig.~\ref{fig:summary}\,d).
In this case, the decoherence effect for an accelerated
observer, hovering at constant height in the gravitational field, is exactly canceled by the potential
that is keeping the particle from falling.
The decoherence effect reported by
Pikovski et~al.~\cite{Pikovski:2015}, therefore, is an effect of \emph{relative acceleration} of particle
and observer (i.\,e. the experimental set-up), rather than a universal decoherence in the presence of
a gravitational field.

\section{Earth bound observers}
It is important to point out the significance of the observer for a quantum mechanical
description. In contrast to classical relativity, where the observer is simply an expression
of the set of coordinates used to describe the physics, in quantum mechanics the bipartite nature
of the theoretical description of experiments must be taken into account: in addition to the
dynamical evolution of a quantum state, this description must also contain the reduction of the state
(either objectively or effectively, regardless of the preferred interpretation of quantum mechanics)
to the eigenstate of some operator upon preparation and measurement.
The most convenient choice for the observer frame is, therefore, generally the rest-frame of the
experimental set-up. Any other observer frame will require an appropriate transformation, not only
of the dynamics but also of the measuring procedure.
In this context, the discussion of decoherence due to a moving detector~\cite{Pang:2016} is
insightful.

The frame attached to an observer moving along a timelike curve $\gamma$ in curved spacetime
is usually described in terms of Fermi normal coordinates.
For nearly-local experiments, i.\,e. observations performed in the vicinity of $\gamma$,
Riemann tensor corrections can be neglected, 
and the analysis of a system in an external gravitational field reduces to the
description of an accelerated observer in flat spacetime~\cite{Misner:1973}.
Specifically, for a radially infalling observer
in the Earth's gravitational field the Fermi normal coordinates
$(ct,x,y,z)$  near a timelike curve $\gamma$ yield the flat spacetime Minkowski metric,
as required by the equivalence principle.
For a shell observer, hovering at constant altitude,
the metric expressed in Fermi normal coordinates reduces to the
Rindler metric~\cite{Poisson:2011,Manasse:1963}
\begin{equation}
\mathrm{d}s^{2}=-\left(1+\frac{gx}{c^{2}}\right)^{2}
c^{2} \mathrm{d}t^{2} + \mathrm{d}x^2 + \mathrm{d}y^2 + \mathrm{d}z^2
+\mathcal{O}(\Delta^{3}),\label{eq:metric}
\end{equation}
where $\Delta$ denotes the spatial distance from $\gamma$, $x$ corresponds
to vertical displacements and $g$ is the acceleration on $\gamma$.

In other words, we have Einstein's presentation of the equivalence principle~\cite{Norton:1985},
according to which the local physics in the gravitational field can be described entirely as
the physics of an accelerated observer in flat spacetime.
The rest of this letter, therefore, will be concerned only with \com\ variables
for Minkowski and Rindler observers.

\section{Relativistic \com\ for inertial observers}
Early works have dealt with the extension of nonrelativistic quantum mechanics to
relativistic single-particle situations where particle creation and annihilation can be 
suppressed~\cite{Foldy:1950,Salpeter:1952}. Newton and Wigner~\cite{Newton:1949} show the
existence of a unique position operator for elementary systems
whose components satisfy the usual commutation relations.
This position operator is frame dependent already in the single-particle case~\cite{Fleming:1965};
a fact that is little surprising since measurements are attached to a spacelike hypersurface in
a \emph{specific} frame.

This framework can be generalized to many particles. Relativistic \com\ variables which satisfy
the canonical commutation relations can then be constructed, starting from the Poincaré
algebra~\cite{Krajcik:1974} (cf. Refs.~\cite{Darwin:1920,Bakamjian:1953} for earlier works).
The ten generators of the Poincaré group (translations $\ppv$,
rotations $\pjv$, boosts $\pkv$, and time translation $\ph$) satisfy the Lie algebra
\begin{equation}\begin{alignedat}{2}\label{eqn:poincare-algebra}
[\pp_i,\pp_j] &= [\pp_i,\ph] \rlap{$ = [\pj_i,\ph] = 0 \,,
\quad\quad (i,j = 1,2,3) $} \\
[\pj_i,\pj_j] &= \mathrm{i}\, \epsilon_{ijk} \pj_k \,, \quad &
[\pj_i,\pp_j] &= \mathrm{i}\, \epsilon_{ijk} \pp_k \,, \\
[\pj_i,\pk_j] &= \mathrm{i}\, \epsilon_{ijk} \pk_k \,, \quad &
[\pk_i,\ph] &= \mathrm{i}\, \pp_i \,, \\
[\pk_i,\pp_j] &= \mathrm{i}\, \delta_{ij} \ph / c^2 \,, \quad &
[\pk_i,\pk_j] &= -\mathrm{i}\, \epsilon_{ijk} \pj_k / c^2 \,.
\end{alignedat}\end{equation}
For a system of $N$ spinless particles with masses $m_\mu$,
the generators can be expressed
in terms of the particle coordinates $\vec{r}_\mu$ and momenta $\vec{p}_\mu$
($\mu = 1,\dots,N$):
\begin{subequations}\label{eqn:poincare-generators}\begin{align}
\ppv &= \sum_{\mu=1}^N \vec{p}_\mu \,, \quad
\pjv = \sum_{\mu=1}^N \vec{r}_\mu \times \vec{p}_\mu \,, \\
\ph &= \sum_{\mu=1}^N T_\mu + U \,, \\
\pkv &= \sum_{\mu=1}^N \left(\frac{1}{2 c^2} \{\vec{r}_\mu,T_\mu\} - t \vec{p}_\mu \right) + \vec{V} \,,
\end{align}\end{subequations}
where $\{\cdot,\cdot\}$ denotes the anti-commutator, $U$ is the interaction potential between the
particles, $\vec{V}$ the ``interaction boost''~\footnote{The inclusion of an interaction term in the
Hamiltonian requires to include a corresponding term in the boost generators for consistency with
the Poincaré algebra~\cite{Foldy:1961}}, and the single-particle kinetic energy is
\begin{equation}\label{eqn:kinetic-energy}
T_\mu = \sqrt{\vec{p}_\mu^2 c^2 + m_\mu^2 c^4} \,.
\end{equation}
The Galilean group algebra can be straightforwardly obtained as the group contraction
$c \to \infty$~\cite{Inonu:1953}, where only the last line of
Eq.~\eqref{eqn:poincare-algebra} changes to
\begin{equation}
[\pk_i,\pp_j] = \mathrm{i}\, \delta_{ij} M \,, \quad
[\pk_i,\pk_j] = 0 \,,
\end{equation}
where $M = \sum m_\mu$ is the total mass.

In the nonrelativistic case, it is natural to introduce the \emph{\com\ coordinates}
$\vec{P} = \sum \vec{p}_\mu$ and $\vec{R} = \sum m_\mu \vec{r}_\mu / M$;
if we also require that the total internal angular momentum be zero,
then the generators of the Galilean group assume a \emph{single particle} form
\begin{equation}\begin{alignedat}{2}\label{eqn:galilei-generators-single-particle}
\ppv &= \vec{P} \,, \quad &
\pjv &= \vec{R} \times \vec{P} \\
\ph &= \frac{\vec{P}^2}{2 M} + H_\text{rel} \,, \quad &
\pkv &= M \vec{R} - t \vec{P} \,.
\end{alignedat}\end{equation}
The internal Hamiltonian $H_\text{rel}$ contains internal kinetic energies
and the interaction potential $U$. The interaction boost $\vec{V}$ is constrained by the shape
of the potential $U$, and its nonrelativistic contribution can be set equal to zero
without loss of generality~\cite{Foldy:1961}.

Krajcik and Foldy~\cite{Krajcik:1974} notice that this single particle form of the generators
can, alternatively, be understood as a \emph{definition} of \com\ variables.
This is the starting point of the relativistic extension of \com\ variables. One requires that the
\emph{Poincaré} generators take a \emph{relativistic} single particle form:
\begin{equation}\begin{alignedat}{2}\label{eqn:poincare-generators-single}
\ppv &= \vec{P} \,, \quad &
\pjv &= \vec{R} \times \vec{P} \\
\ph &= \sqrt{\vec{P}^2 c^2 + H_\text{rel}^2} \,, \quad &
\pkv &= \frac{1}{2 c^2} \{\vec{R},\ph\} - t \vec{P} \,.
\end{alignedat}\end{equation}
The \com\ variables can then be constructed in a perturbative way order by order in $1/c^2$.
In this construction, the Hamiltonian describing the \com\ motion in Minkowski spacetime---or the
motion of a free falling particle as seen by a free falling observer---follows immediately from
the time translation Poincaré group generator:
\begin{equation}\begin{split}\label{eqn:h-mink}
H^\text{Mink.}_\text{\com} = \sqrt{\vec{P}^2 c^2 + H_\text{rel}^2}.
\end{split}\end{equation}
It is then straightforward to obtain the lowest order relativistic correction by expanding to order $1/c^2$:
\begin{equation}\begin{split}\label{eqn:h-mink_approx}
H^\text{Mink.}_\text{\com} \approx M c^2 + \frac{\vec{P}^2}{2 M} + H_\text{rel}^{(0)} \\
+\frac{1}{c^2} \left( -\frac{\vec{P}^4}{8 M^3} + H_\text{rel}^{(1)} - \frac{\vec{P}^2 H_\text{rel}^{(0)}}{2 M^2} \right) \,.
\end{split}\end{equation}
In the first line, this Hamiltonian consists of the rest mass energy and the nonrelativistic kinetic
and internal energies. The second line contains the relativistic corrections, both for the kinetic
energy and the internal interactions. The last term couples the (nonrelativistic) \com\ kinetic energy
to the (nonrelativistic) internal energy.
The explicit form of the \com\ coordinates and of the internal Hamiltonian can be found in Ref.~\cite{Krajcik:1974}.

\begin{figure}
\includegraphics[scale=.5]{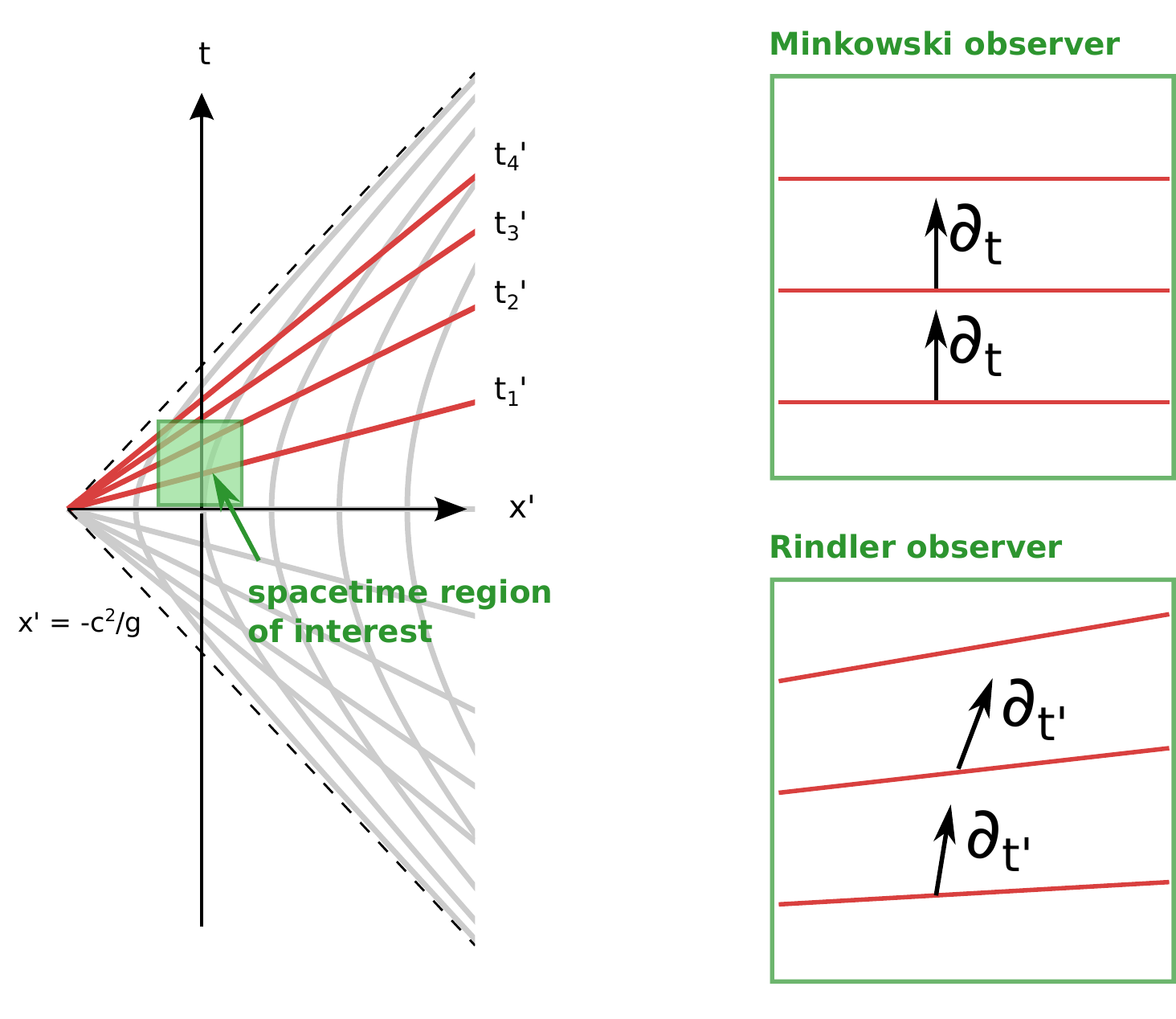} 
\caption{Transformation from free falling Minkowski observer to accelerated Rindler observer.
In the spacetime diagram on the left hand side, the simultaneity hypersurfaces for the
Rindler observer are depicted by the red lines, with the dotted line representing the Rindler
horizon. We also include the trajectories given by hyperbolic rotations.
The spacetime region of interest is highlighted in green and the magnified drawing on the
right hand side compares the simultaneity hypersurfaces for the Rindler and Minkowski observers
and depicts the corresponding time-evolution Killing vectors.
\label{fig:rindler}}
\end{figure}

\section{Relativistic \com\ for accelerated observers}
So far, only inertial systems have been considered.
Let us now discuss the dynamics of the same free falling composite particle, but from the perspective of an accelerated
observer. In general, this can be considered as a noninertial transformation from Minkowski to Rindler
coordinates (see Fig.~\ref{fig:rindler}). We can construct the Hamiltonian operator for the Rindler
observer by considering a family of inertial frames instantaneously at rest with the Rindler observer.
We express the Rindler time evolution Killing vector $ \partial_{t'}$ in the coordinates of the
instantaneous inertial frame, and use the correspondence
between Killing vectors and the generators of the Poincar\'{e} group in inertial frames.
In this construction, the Hamiltonian describing the \com\ motion in Rindler spacetime---or
the motion of a free falling particle as seen by a suspended observer on Earth---is given by
(cf. appendix~\ref{sec:kv} for a step by step construction):
\begin{equation}\label{rindlerH}
H^\text{Rindler}_\text{\com}=H^\text{Mink.}_\text{\com}+ \frac{g}{2 c^2} \{ X,H^\text{Mink.}_\text{\com} \}\,,
\end{equation}
where these operators are defined on the Rindler equal-time hypersurfaces. 
Note that this is the well-known Rindler Hamiltonian \cite{unruh1984acceleration,unruh1976notes} rewritten within Foldy's formalism. Again,
the lowest order relativistic correction is obtained by expanding to order $1/c^2$:
\begin{subequations}\label{eqn:hrindler}\begin{align}
H^\text{Rindler}_\text{\com} &= M c^2 + H^{(0)} + \frac{1}{c^2} H^{(1)} \\
H^{(0)} &= \frac{\vec{P}^2}{2 M} + H_\text{rel}^{(0)} + M\,g\,X +  U_\text{ext}^{(0)} \\
H^{(1)} &= -\frac{\vec{P}^4}{8 M^3} + H_\text{rel}^{(1)} - \frac{\vec{P}^2 H_\text{rel}^{(0)}}{2 M^2}\nnl
&\bleq + \frac{g}{4\,M} \{X, \vec{P}^2\} + H_\text{rel}^{(0)}\,g \,X +  U_\text{ext}^{(1)} \,,
\end{align}\end{subequations}
In addition to the terms present in the Minkowski Hamiltonian~\eqref{eqn:h-mink_approx}, at the nonrelativistic level one obtains the homogeneous potential $M g X$, as expected. The relativistic corrections to this gravitational term consist of a pure \com\ term, coupling the \com\ position to the \com\ kinetic energy, as well as the coupling of \com\ position to the internal Hamiltonian. The latter is the term of interest when discussing potential decoherence effects~\cite{Pikovski:2015}. We also included an external interaction $U_\text{ext}$.

In Refs.~\cite{Laemmerzahl:1995,Pikovski:2015} the Hamiltonian of a particle in the
gravitational field of the Earth was derived as the limit of Schwarzschild spacetime
for radii large compared to the Schwarzschild radius. The result differs in two
respects from ours:
an additional term $\frac{1}{2}M g^2 X^2$, expressing curvature corrections
which were neglected in our limit from Fermi normal coordinates to Rindler coordinates,
as well as a pre-factor of 3 in the term proportional to $\{X, \vec{P}^2\}$, 
which is easily understood comparing the Schwarzschild metric to the Rindler
metric~\footnote{While for the Rindler metric only the tt-component contains acceleration
terms, in the Schwarzschild metric there are gravitational terms also in the spatial part.}.
However, recalling the above discussion of the role of the observer, it is important to
keep in mind that the description in Schwarzschild coordinates is that of an observer at
infinite radial distance. This is not an issue in any classical situation, where the
choice of coordinates is only a matter of convenience, but in the quantum scenario
direct predictions for experimental outcomes must be made in a frame attached to the
laboratory.

\section{Supporting external potential}
In general, the external interaction $U_\text{ext}$ can of course be of any form, depending on the
experimental situation. Notably, it can also depend on time, \com\ momentum, or internal degrees
of freedom of the particle.
We now want to discuss the requirements on this potential in order to
support the particle and keep it from falling (i.\,e. be at rest with respect to the observer).
A more detailed discussion is provided in the appendix~\ref{sec:sup}.

Note that a potential $U_\text{ext}(X)$, which would have no effect on the coupling of \com\
and internal dynamics~\cite{Shariati:2016,Pikovski:2015a}, cannot keep the particle from falling
in a relativistic situation. For a classical system the argument is quite straightforward.
In order to keep the particle at rest, we
need to require a static momentum. With Hamilton's equations of motion,
$\dot{\vec P} = -\partial H^\text{Rindler}_\text{\com} / \partial \vec R = 0$, this implies
that the external interaction must exactly cancel all $g$-dependent terms, including the
coupling of \com\ position to internal Hamiltonian and, therefore, the potential cannot depend
on the \com\ coordinate $X$ only.

The necessary condition for a \emph{quantum} particle in order to be at rest in the observer's
frame can be expressed as the condition that the position expectation value is not accelerated:
$\D^2 \langle X \rangle / \D t^2 = 0$. This only requires the classical behavior to mimic the evolution
in Minkowski spacetime. It does not reinforce any further restrictions on the quantum dynamics.
Using the Ehrenfest theorem, at the nonrelativistic level
this yields the expected result: $U_\text{ext}^{(0)} = -M g X$.

At the order $1/c^2$ the condition for the acceleration of $\langle X \rangle$ to vanish is
\begin{equation}\label{eqn:vanishingcondition}\begin{split}
\langle[P_x,{U}_\text{ext}^{(1)}]\rangle 
-\frac{\mathrm{i}}{2\hbar} \langle [[X,{U}_\text{ext}^{(1)}],\vec{P}^2] \rangle
\\= \mathrm{i} \hbar g \left( \langle H_\text{rel}^{(0)} \rangle
+ \frac{\langle \vec{P}^2 \rangle - 2 \langle P_x^2 \rangle}{2 M} \right)
\end{split}\end{equation}
Although this condition could be satisfied by a time-dependent potential which is tuned to the 
respective expectation values, a more realistic model for physical interactions is an operator-valued
$U_\text{ext}^{(1)}$, depending on both position and momentum~\footnote{Such a momentum dependent interaction
follows, for instance, for a Klein-Gordon field minimally coupled to an electric field.}.
In this case, the condition~\eqref{eqn:vanishingcondition} yields the interaction
\begin{equation}
U_\text{ext}^{(1)} = -H_\text{rel}^{(0)} g X -\frac{g}{4 M} \{X , \vec{P}^2\} \,.
\end{equation}
Unsurprisingly, this interaction exactly cancels all acceleration terms
in the Hamiltonian~\eqref{eqn:hrindler}.

\section{Conclusion}
The term $c^{-2} H_\text{rel}^{(0)} g X$ in the Hamiltonian~\eqref{eqn:hrindler} which couples the internal energy
of a quantum system to its position through acceleration has been the source for the
prediction of
decoherence~\cite{Pikovski:2015}. Our discussion from the perspective of the Poincaré group and
the inclusion of the external interaction clarifies the meaning of this result and answers some of the
concerns about whether it is in contradiction to the equivalence principle.

The predicted decoherence effect clearly exists for a free falling particle observed in an
accelerated laboratory (or a laboratory on Earth) as depicted in Fig.~\ref{fig:summary}\,(b). 
Experimentally, this is the situation one finds for example in interferometry experiments 
with atoms~\cite{kovachy2015quantum} and molecules~\cite{Arndt:2014,Eibenberger:2013}.
From the perspective of the equivalence principle,
it is important to point out that it is the detector and not the particle that is
accelerated in this situation.
In this regard, Bonder et~al.~\cite{Bonder:2016} rightfully remark that the situation analysed by
Pikovski et~al.~\cite{Pikovski:2015} is equivalent to that of an accelerated observer studying
free, isolated systems without gravity; however, their conclusion that ``such scenarios cannot
lead to decoherence as, without gravity, there is nothing to cause it'' is incorrect.
(Of course, the equivalence principle allows to describe the particle evolution both from the perspective
of the accelerated Rindler observer moving with the detector and from the perspective of the free
falling Minkowski frame moving with the particle.) Although there is no gravity in the latter case,
decoherence still results from the \emph{accelerated} detector~\cite{Pang:2016}.
On the other hand, a free falling detector measuring the very same particle will, of course, not see any decoherence.

For a particle which is held at constant position in the laboratory frame, as in
Fig.~\ref{fig:summary}\,(d), we learn that the
interaction required to keep the particle from falling cannot be ignored in the discussion.
Quite to the contrary, this interaction will generally cancel the coupling terms that are
supposed to lead to decoherence.

The requirement of canonical commutation relations for the \com\ variables---which is the only
choice that can easily be reconciled with the principles of Quantum Mechanics---leads to the
definition adopted from Krajcik and Foldy~\cite{Krajcik:1974}.
The resulting Hamiltonians~\eqref{eqn:h-mink_approx} and~\eqref{eqn:hrindler} should be understood as
generators of the dynamics of the \com\ coordinate \emph{in this particular choice}.
It must be stressed that the \com\ coordinate found in this way is \emph{frame dependent},
and a distinguished role is given to the instantaneous rest frame of the detector.
Although this seems to be the most plausible extension of the principles of Quantum Mechanics to
relativistic situations, the ultimate decision about whether or not it is correct must be made
by experiment.

Note that acceleration actually couples the \com\ position to the \emph{full}
Hamiltonian, as is evident from Eq.~\eqref{rindlerH}.
Even in absence of acceleration, or when an external potential cancels all contributions
of the observer's acceleration,
the remaining special relativistic correction proportional to $\vec P^2 H_\text{rel}^{(0)}$ still couples the
\com\ dynamics to the internal Hamiltonian.

\begin{acknowledgments}
The authors thank D.~Giulini, D.~Sudarsky, and S.~Bacchi for insightful discussions, and gratefully
acknowledge funding and support from {\small{INFN}} and the University of Trieste (FRA 2016).
A.G. acknowledges funding from the German Research Foundation (DFG).
\end{acknowledgments}

\appendix

\section{Step by step construction of Rindler Hamiltonian}\label{sec:kv}
Let us first review how to obtain the Rindler space Killing vectors in the coordinates
of the instantaneous inertial frame \cite{rindler1969essential}.
Consider a family of inertial frames instantaneously at rest with
the Rindler observer: specifically, consider the reference frame
$S(\bar{t}')$ at the Rindler time $\bar{t}'$. The transformation
from the Rindler coordinates $x^{\mu'}=(ct',x')$ to the $S(\bar{t})$
coordinates $x^{\mu}=(cT,X)$ is given by:

\begin{alignat}{1}
cT & =(x'+\frac{c^{2}}{g})\,\text{sinh}(\frac{g(t'-\bar{t}')}{c}),\\
X & =(x'+\frac{c^{2}}{g})\,\text{cosh}(\frac{g(t'-\bar{t}')}{c})-\frac{c^{2}}{g}.
\end{alignat}
We omit the other two coordinates $Y=y'$ and $Z=z'$ in the discussion.
The inverse transformation is given by:

\begin{alignat}{1}
ct' & =\frac{c^{2}}{g}\text{tanh}^{-1}(\frac{cT}{X+\frac{c^{2}}{g}})+c\bar{t}',\label{eq:tr1i}\\
x' & =\sqrt{(X+\frac{c^{2}}{g})^{2}-c^{2}T^{2}}-\frac{c^{2}}{g}.\label{eq:tr2i}
\end{alignat}

We now consider the Killing vectors $\frac{1}{c}\partial_{t'}$, $\frac{1}{c}\partial_{T}$,
$\partial_{X}$: the Minkowski metric, expressed in terms of the Rindler
coordinates or the $S(\bar{t}')$ coordinates, does not depend on
$t'$ or $T$, $X$, respectively. In particular, we will rewrite
$(\frac{1}{c}\partial_{t'})^{\mu'}=(1,0)$ in terms of $(\frac{1}{c}\partial_{T})^{\mu}=(1,0)$
and $(\partial_{X})^{\mu}=(0,1)$ at the time $t'=\bar{t}'$:

\begin{equation}
v^{\mu}=\frac{\partial x^{\mu}}{\partial x^{\mu'}}\vert_{t'=\bar{t}'}\left(\frac{1}{c}\partial_{t'}\right)^{\mu'}\label{eq:vector}
\end{equation}
To this end we calculate the transformation matrix $\frac{\partial x^{\mu'}}{\partial x^{\mu}}\vert_{t'=\bar{t}'}$ from
Eqs. (\ref{eq:tr1i}), (\ref{eq:tr2i}), which can be easily inverted
to give:

\begin{equation}
\frac{\partial x^{\mu}}{\partial x^{\mu'}}\vert_{t'=\bar{t}'}=\left[\begin{array}{cc}
1+gX/c^{2} & 0\\
0 & \text{sign}(1+gX/c^{2})
\end{array}\right]\label{eq:tmi}
\end{equation}
where $\text{sign}(a)=a/\abs{a}$ is the sign function.

It is then straightforward,
using Eqs. (\ref{eq:vector}), (\ref{eq:tmi}), to obtain $v^{\mu}=(1+gX/c^{2},0)$,
i.\,e.

\begin{equation}\label{conversion}
\partial_{t'}=(1+g/c^{2}X)\partial_{T}.
\end{equation}

\subsection{From Killing vectors to Hilbert space operators}
Now let us construct the Hamiltonian operator $\hat{H}$ that generates the time evolution on the Hilbert space $H$ for inertial and non-inertial observers. As we will discuss in detail, the operator $\hat{H}$ depends on the type of motion, i.\,e. inertial or non-inertial, in general can change over time, and is also hypersurface dependent. This is not surprising, but rather expected within the chosen framework~\cite{Krajcik:1974}: the experimental apparatus, modeled by operators on Hilbert space, performs measurements on equal-time hypersurfaces.

In inertial reference frames, the scheme of the derivation is to
map the time evolution Killing vector (corresponding to the
Poincar\'{e} algebra time evolution generator) on an equal-time
hypersurface to an operator on Hilbert space: this operator
generates the infinitesimal time evolution associated to the
specific motion and to the specific hypersurface. To construct
this map, i.e. the representation of the Poincar\'{e} algebra on
the Hilbert space $H$, we consider an inertial reference frame with
coordinates $(cT, X_1, X_2, X_3)$. The isomorphism $\xi$ between
elements of Poincar\'{e} algebra $\mathcal{A}$ and the Killing vectors  $V$ (at time $T=0$) is given by
\begin{align}
\mathcal{H}   &\mapsto \partial_T, \label{KillingH}\\
\mathcal{K}_i &\mapsto  T\frac{\partial}{\partial_{X_i}}\big|_{T=0} + X_i \frac{\partial}{c^2\partial T}\big|_{T=0} = X_i \frac{\partial}{c^2\partial T}\big|_{T=0},\\
\mathcal{P}_i &\mapsto  \partial_{X_i},\\ 
\mathcal{J}_i &\mapsto  \epsilon_{ijk} X_j\partial_{X_k} \,
\end{align}
and the Lie algebra representation $\gamma$ on the Hilbert space $H$ (at time $T=0$) is given by:
\begin{align}
\mathcal{H}   &\mapsto \hat{H}, \label{opH}\\
\mathcal{K}_i &\mapsto \hat{K}_i\big|_{T=0}, \label{opK}\\
\mathcal{P}_i &\mapsto  \partial_{X_i}, \\
\mathcal{J}_i &\mapsto  \epsilon_{ijk} X_j\partial_{X_k}. \label{opJ}
\end{align}
This establishes the linear map (homomorphism) $\phi=\gamma \circ \xi^{-1}$ between Killing vectors $V$ and Hilbert space operators $A(H)$ in an inertial frame (at time $T=0$). We have the following picture
\begin{displaymath}
    \xymatrix{
    	\mathcal{A} \ar[r]^{\xi} \ar[d]^{\gamma} & V \ar[ld]^{\phi} \\
		A(H)},
\end{displaymath}
where $\mathcal{A}$, $V$ and $A(H)$ denote the Poincar\'{e} algebra, the Killing vectors and the algebra of Hilbert space operators, respectively. 

We first discuss the situation for the Minkowski observer in inertial motion, where the coordinates are denoted by $(ct,x_1,x_2,x_3)$.  Specifically, at time $t=\bar{t}$ we consider a new inertial frame with coordinates $(cT,X_1,X_2,X_3)$ such that the $T=0$ hypersurface corresponds to the $t=\bar{t}$ hypersurface, and we set $x_i=X_i$. Using this new inertial frame one then establishes that the generator of time evolution on the equal-time hypersurface $T=0$ is the operator $\hat{H}$, i.e. we use the map between Killing vectors and Hilbert space operators:
\begin{equation}
\phi: \partial_T \mapsto \hat{H}.
\end{equation}
We define the  Minkowski Hamiltonian $\hat{H}^{\text{Mink.}}$ on
the $t=\bar{t}$ hypersurface to be given by the operator $\hat{H}$.
We repeat this construction on each equal-time hypersurface of the
inertial (Minkowski) observer by varying $\bar{t}$: in this way we
define $\hat{H}^{\text{Mink.}}$ for all $t$.
Specifically, the Minkowski Hamiltonian in \com\ coordinates is
given by 
\begin{equation}
\hat{H}^\text{Mink.}_\text{\com} = \sqrt{\hat{\vec{P}}^2 c^2 + \hat{H}_\text{rel}^2},
\end{equation}
where we have inserted the time-evolution generator expressed in \com\ coordinates (see Eq.~(7) from the main text).

We next discuss time evolution for the non-inertial (Rindler)
observer, where the coordinates are denoted by $(ct',x'_1,x'_2,x'_3)$.
We will define the Hamiltonian $\hat{H}^\text{Rindler}$ for the
Rindler observer exploiting the map between Killing vectors and
Hilbert space operators in inertial frames (see Eqs.\eqref{KillingH} -\eqref{opJ}). Specifically, at time $t'=\bar{t}'$ we consider an inertial frame with coordinates $(cT,X_1,X_2,X_3)$ such that the $T=0$ hypersurface corresponds to the $t'=\bar{t}'$ hypersurface, and we set $x'_i=X_i$. We rewrite the time evolution Killing vector $\partial_{t'}$ in the coordinates of the inertial frame instantaneously at rest with the Rindler observer on the $T=0$ hypersurface (see Eq. \eqref{conversion}), i.e. $\partial_{t'}=\partial_{T}+g\frac{X}{c^{2}}\partial_{T}$. From this point onwards, the derivation mirrors the derivation of the previous paragraph for the Minkowski observer. Using the linearity of the map $\phi$ from Killing vectors to Hilbert space operators (see Eqs.\eqref{KillingH} -\eqref{opJ}) we obtain the following operator (at time $t'=\bar{t}'$): 
\begin{equation}
\phi: \partial_{T}+g\frac{X_i}{c^{2}}\partial_{T}\big|_{T=0}
\mapsto \hat{H}+g\hat{K}_i\big|_{T=0}.
\end{equation}
We define the  Rindler Hamiltonian $\hat{H}^{\text{Rindler}}$ at
time $t=\bar{t}'$ ($T=0$) to be given by the operator:
\begin{equation}\label{eq:Rindler_H}
\hat{H}^{\text{Rindler}}=\hat{H}+g\hat{K}_i\big|_{T=0}.
\end{equation}
We repeat this construction on each equal-time hypersurface
of the non-inertial (Rindler) observer by varying $\bar{t}'$:
in this way we define $\hat{H}^{\text{Rindler}}$ for all $t'$.
We now consider  accelerated motion along the  $x'_1$  axis.
From Eq.~\eqref{eq:Rindler_H} it is straightforward to write the
Rindler Hamiltonian in \com\ coordinates:
\begin{equation}
\hat{H}^\text{Rindler}_\text{\com}=\hat{H}^\text{Mink.}_\text{\com}+ \frac{g}{2 c^2} \{ \hat{X},\hat{H}^\text{Mink.}_\text{\com} \}\,,
\end{equation}
where we have inserted the time-evolution and boost generators expressed in \com\ coordinates (see Eq.~(7) from the main text) and $\hat{H}^\text{Mink.}_\text{\com} = \sqrt{\hat{\vec{P}}^2 c^2 + \hat{H}_\text{rel}^2}$.

\section{Restrictions on supporting potentials}\label{sec:sup}
For the \com\ motion in the local coordinates of an accelerated
observer we found the Hamiltonian
\begin{subequations}\begin{align}
H^\text{Rindler}_{\com} &= H^\text{Mink.} + \frac{g}{2c^2} \{X,H^\text{Mink.}\} + U_\text{ext} \label{eqn:HRindler} \\
H^\text{Mink.} &= M c^2 + H^{(0)} + \frac{1}{c^2} H^{(1)} \\
H^{(0)} &= \frac{\vec{P}^2}{2M} + H_\text{rel}^{(0)} \label{eqn:hamzero} \\
H^{(1)} &= -\frac{\vec{P}^4}{8M^3} + H_\text{rel}^{(1)} - \frac{\vec{P}^2 H_\text{rel}^{(0)}}{2 M^2}
\end{align}\end{subequations}
We will also need the commutation relations
\begin{subequations}\begin{align}
[X,H^{(0)}] &= \frac{\mathrm{i}\hbar}{M} P_x \\
[X,\{X,H^{(0)}\}] &=
\frac{\mathrm{i}\hbar}{M} \,\{X,P_x\} \\
[[X,H^{(0)}],\{X,H^{(0)}\}] &=
\frac{2\hbar^2}{M} H^{(0)} \\
[[X,\{X,H^{(0)}\}],H^{(0)}] &=
-\frac{2\hbar^2}{M^2} P_x^2 \\
[[X,\{X,H^{(0)}\}],X] &=
\frac{2\hbar^2}{M} X 
\end{align}\end{subequations}
With the Ehrenfest theorem,
\begin{equation}
\frac{\D}{\D T} \langle A \rangle = -\frac{\mathrm{i}}{\hbar} \langle [A,H] \rangle
+ \left\langle \frac{\partial A}{\partial T} \right\rangle \,,
\end{equation}
one obtains the equations of motion for the \com\ expectation value:
\begin{equation}
\frac{\D^2}{\D T^2} \langle X \rangle = -\frac{1}{\hbar^2} \left\langle [[X,H],H] \right\rangle
- \frac{\mathrm{i}}{\hbar} \langle[X,\frac{\partial}{\partial T}U_\text{ext}]\rangle \,,
\end{equation}
where, for now, we allow for the external potential to be time dependent.
Inserting the Rindler Hamiltonian~\eqref{eqn:HRindler}, we obtain:
\begin{align}
\frac{\D^2\langle X \rangle}{\D T^2} 
&= -\frac{g}{2\hbar^2 c^2} \left\langle [[X,H^\text{Mink.}],\{X,H^\text{Mink.}\}] \right\rangle
-\frac{1}{\hbar^2} \left\langle [[X,H^\text{Mink.}],U_\text{ext}] \right\rangle\nnl
&\bleq
-\frac{g}{2\hbar^2 c^2} \left\langle [[X,\{X,H^\text{Mink.}\}],H^\text{Mink.}] \right\rangle 
-\frac{g^2}{4\hbar^2 c^4} \left\langle [[X,\{X,H^\text{Mink.}\}],\{X,H^\text{Mink.}\}] \right\rangle\nnl
&\bleq
-\frac{g}{2\hbar^2 c^2} \left\langle [[X,\{X,H^\text{Mink.}\}],U_\text{ext}] \right\rangle
-\frac{1}{\hbar^2} \left\langle [[X,U_\text{ext}],H^\text{Mink.}] \right\rangle \nnl
&\bleq
-\frac{g}{2\hbar^2 c^2} \left\langle [[X,U_\text{ext}],\{X,H^\text{Mink.}\}] \right\rangle 
-\frac{1}{\hbar^2} \left\langle [[X,U_\text{ext}],U_\text{ext}] \right\rangle
- \frac{\mathrm{i}}{\hbar} \langle[X,\frac{\partial}{\partial T}U_\text{ext}]\rangle
\end{align}

\newcommand{\uo}{U_\text{ext}^{(0)}}
\newcommand{\ue}{U_\text{ext}^{(1)}}

\subsection{Classical, time dependent potential}\label{app:classical}
Let us first discuss how the particle can be kept from falling, i.\,e. be co-accelerated with
the Rindler observer, by a classical potential which is assumed to be a function of $X$ and $T$:
\begin{equation}
U_\text{ext}(T,X) = \uo(T,X) + \frac{1}{c^2} \ue(T,X) \,.
\end{equation}
Hence, all commutators of $U_\text{ext}$ with $X$ are zero and
at lowest order $(1/c)^0$ we get
\begin{align}\label{eqn:lowestorderacc}
\left.\frac{\D^2\langle X \rangle}{\D T^2}\right|^{\mathcal{O}(c^0)}
&= -\frac{g M}{\hbar^2} \left\langle [[X,H^{(0)}],X] \right\rangle
-\frac{1}{\hbar^2} \left\langle [[X,H^{(0)}],\uo] \right\rangle \nnl
&= -\frac{\mathrm{i} g}{\hbar} \left\langle [P_x,X] \right\rangle
-\frac{\mathrm{i}}{\hbar M} \left\langle [P_x,\uo] \right\rangle 
= -\frac{\mathrm{i}}{\hbar M} \left\langle [P_x,M g X + \uo] \right\rangle \,,
\end{align}
which vanishes for the choice $\uo = - M g X$, as expected.
With this result, the first order correction becomes
\begin{subequations}\begin{align}
\left.c^2\,\frac{\D^2\langle X \rangle}{\D T^2}\right|^{\mathcal{O}(c^{-2})}
&=
-\frac{g}{2\hbar^2} \left\langle [[X,H^{(0)}],\{X,H^{(0)}\}] \right\rangle
-\frac{1}{\hbar^2} \left\langle [[X,H^{(0)}],\ue] \right\rangle \nnl
&\bleq
-\frac{g}{2\hbar^2} \left\langle [[X,\{X,H^{(0)}\}],H^{(0)}] \right\rangle \\
&=
-\frac{g}{M} \left\langle H^{(0)} \right\rangle
-\frac{\mathrm{i}}{\hbar M} \left\langle [P_x,\ue] \right\rangle
+\frac{g}{M^2} \left\langle P_x^2 \right\rangle \,.
\end{align}\end{subequations}
Requiring this acceleration of the particle to vanish (for the Rindler observer),
using the definition~\eqref{eqn:hamzero} of $H^{(0)}$, we obtain the condition
\begin{equation}\label{eqn:vanishingcondition-class}
\langle[P_x,\ue]\rangle = \mathrm{i} \hbar g \left( \langle H_\text{rel}^{(0)} \rangle
+ \frac{\langle \vec{P}^2 \rangle - 2 \langle P_x^2 \rangle}{2 M} \right) \,,
\end{equation}
which is satisfied for the potential
\begin{equation}\label{eqn:uext-class}
\ue(T) = -\left( \langle H_\text{rel}^{(0)} \rangle + \frac{\langle \vec{P}^2 \rangle - 2 \langle P_x^2 \rangle}{2 M} \right) g X \,.
\end{equation}
Therefore, the potential keeping the particle from falling to order $1/c^2$ will generally be
\emph{time dependent}, and must be tuned to match the expectation values of momentum squared and
internal energy.

This potential can be a reasonable choice in order to provide an effective description of the
classical interventions of the experimenter, e.\,g. externally tuning a trapping potential in
such a way that a self-consistent behavior is achieved.
However, as a \emph{fundamental} equation, describing external interactions of the quantum system,
the resulting \emph{nonlinear} Schrödinger equation would lead
to superluminal signaling ~\cite{Gisin:1989}.
To this end, the Schrödinger equation with the potential~\eqref{eqn:uext-class} is not
a general evolution equation, but only applies to a specific choice of initial \com\
state and external potential.

\newcommand{\uoh}{\hat{U}_\text{ext}^{(0)}}
\newcommand{\ux}{\hat{\widetilde{U}}_\text{ext}^{(1)}}
\newcommand{\up}{\hat{\widetilde{U}}_{\vec{P}}^{(1)}}

\newcommand{\uq}{\hat{U}_\text{ext}^{(1)}}
\subsection{Quantum interaction with external systems}
A realistic model for a relativistic interaction keeping the particle from falling---the forces
due to the electromagnetic interaction with the atoms of a table, for instance, on which the
particle is at rest---will \emph{not} be of the form~\eqref{eqn:uext-class}, as relativistic
interactions generally do not depend on position only but also on momentum.
As an example, take the Darwin Lagrangian~\cite{Darwin:1920}, or a Klein-Gordon field minimally
coupled to an electric potential $\phi$, where relativistic corrections to the Schrödinger equation
resulting from a $1/c^2$ expansion~\cite{Kiefer:1991} will contain momentum dependent terms
proportional to $\phi \vec P^2$.

Hence, for a more physical model for the interaction, we ask for a $\hat{U}_\text{ext}$ which is an
operator valued function, not only of the
position operator but also of momentum. Appealing to the situation where an equilibrium of forces
is achieved, e.\,g. for the electromagnetic interaction with the atoms of, say, a table, we assume
that this function has no explicit time dependence, and make the ansatz
\begin{equation}
\hat{U}_\text{ext} = \uoh + \frac{1}{c^2} \uq \,,
\end{equation}
where we require that $\uq$ can depend on $\vec P$
and, therefore, does not commute with $X$, while $\uoh$ does commute with $X$.
At lowest order $(1/c)^0$ we then get exactly the same expression~\eqref{eqn:lowestorderacc}
as in subsection~\ref{app:classical}, yielding again $\uo = - M g X$.
The first order correction now becomes
\begin{subequations}\begin{align}
\left.c^2\,\frac{\D^2\langle X \rangle}{\D T^2}\right|^{\mathcal{O}(c^{-2})}
&=
-\frac{g}{2\hbar^2} \left\langle [[X,H^{(0)}],\{X,H^{(0)}\}] \right\rangle
-\frac{1}{\hbar^2} \left\langle [[X,H^{(0)}],\uq] \right\rangle \nnl &\bleq
-\frac{g}{2\hbar^2} \left\langle [[X,\{X,H^{(0)}\}],H^{(0)}] \right\rangle
-\frac{1}{\hbar^2} \left\langle [[X,\uq],H^{(0)}] \right\rangle \\
&=
-\frac{g}{M} \left\langle H^{(0)} \right\rangle
-\frac{\mathrm{i}}{\hbar M} \left\langle [P_x,\uq] \right\rangle 
+\frac{g}{M^2} \left\langle P_x^2 \right\rangle\nnl
&\bleq
-\frac{1}{\hbar^2} \left\langle [[X,\uq],H^{(0)}] \right\rangle \,.
\end{align}\end{subequations}
Rearranging this expression to collect the terms that depend and do not depend on
the external potential, respectively, and requiring the acceleration to vanish
(for the Rindler observer), as before, then yields
\begin{equation}
\langle[P_x,\uq]\rangle
-\frac{\mathrm{i} M}{\hbar} \langle [[X,\uq],H^{(0)}] \rangle
= \mathrm{i} \hbar g \left( \langle H^{(0)} \rangle
- \frac{\langle P_x^2 \rangle}{M} \right) \,.
\end{equation}
Using the definition~\eqref{eqn:hamzero} of $H^{(0)}$, this further simplifies:
\begin{equation}\label{eqn:vanishingcondition-quant}
\langle[P_x,\uq]\rangle 
-\frac{\mathrm{i}}{2\hbar} \langle [[X,\uq],\vec{P}^2] \rangle
= \mathrm{i} \hbar g \left( \langle H_\text{rel}^{(0)} \rangle
+ \frac{\langle \vec{P}^2 \rangle - 2 \langle P_x^2 \rangle}{2 M} \right) \,.
\end{equation}
In order to satisfy this relation, $\uq$ can contain arbitrary position independent
terms ($\vec P$, $\vec P^2$, $\vec P^3$, ...), but on top of that only terms linear in $X$
not higher than second order in $\vec P$ are possible. Considering the commutators
\begin{alignat}{2}
[P_x,X] &= -\mathrm{i} \hbar \,,&\quad\quad
[[X,X],\vec{P}^2] &= 0 \,,\\
[P_x,X P_x] &= -\mathrm{i} \hbar P_x \,,&\quad\quad
[[X,X P_x],\vec{P}^2] &= -\hbar^2 P_x \,,\\
[P_x,X \vec{P}^2] &= -\mathrm{i} \hbar \vec{P}^2 \,,&\quad\quad
[[X,X \vec{P}^2],\vec{P}^2] &= -4\hbar^2 P_x^2 \,,\\
[P_x,\vec{P}^2 X] &= -\mathrm{i} \hbar \vec{P}^2 \,,&\quad\quad
[[X,\vec{P}^2 X],\vec{P}^2] &= -4\hbar^2 P_x^2 \,,
\end{alignat}
we see that the no-acceleration condition is satisfied for
\begin{equation}
\uq = -H_\text{rel}^{(0)} g X -\frac{g}{2 M} (\alpha X \vec{P}^2 + \beta \vec{P}^2 X) \,,
\end{equation}
with $\alpha + \beta = 1$.
Demanding that the $X \vec{P}^2$ and $\vec{P}^2 X$ terms enter symmetrically,
we obtain
\begin{equation}
\uq = -H_\text{rel}^{(0)} g X -\frac{g}{4 M} \{X , \vec{P}^2\} \,.
\end{equation}

\end{document}